\documentclass{PoS}

\newcommand{\be}{\begin{eqnarray}}
\newcommand{\ee}{\end{eqnarray}}
\newcommand{\ba}{\begin{array}}
\newcommand{\ea}{\end{array}}
\newcommand{\bi}{\begin{itemize}}
\newcommand{\ei}{\end{itemize}}

\title{Toward modelization  of quark and gluon transversity  generalized parton distributions  }

\ShortTitle{Towards modelization  of quark and gluon transversity  GPDs }

\author{B.~Pire \\
          CPhT, \'{E}cole Polytechnique, CNRS,  91128 Palaiseau, France
}
\author{
\speaker{K.~Semenov-Tian-Shansky}\\
        IFPA, D\'{e}partement AGO,  Universit\'{e} de  Li\`{e}ge, 4000 Li\`{e}ge,  Belgium \\
        E-mail: \email{ksemenov@ulg.ac.be}}

\author{L.~Szymanowski \\
          National Centre for Nuclear Research (NCBJ), Warsaw, Poland
}

\author{S.~Wallon \\
          LPT,   Universit\'{e} Paris-Sud, CNRS, 91405 Orsay, France \\
           UPMC, Universit\'{e} Paris 06, Facult\'{e} de Physique, 4 place Jussieu, 75252 Paris, France
}

\abstract{Quark and gluon helicity flip generalized parton distributions (GPDs) encode the information on the
nucleon structure in the transversity sector. In order to build a theoretically consistent phenomenological parametrization
for these hadronic matrix element within the framework of the dual parametrization of GPDs (or with the equivalent
approach of the SO$(3)$ partial waves (PW) expansion with the Mellin-Barnes integral techniques) we establish the set of
combinations of parton helicity flip GPDs suitable for the expansion in the cross channel SO$(3)$ PWs. 
}

\FullConference{XXII. International Workshop on Deep-Inelastic Scattering and Related Subjects,\\
		28 April - 2 May 2014\\
		Warsaw, Poland}

\begin{document}

\section{Introduction and preliminaries}
The transversity partonic structure of hadrons constitutes a longstanding challenge for
both theoretical and experimental studies
\cite{Barone:2001sp}.
The appealing feature of the transversity dependent sector is that due to the chiral-odd property
of transversity quark distributions it turns out to be possible to clearly separate quark and gluonic contents.
In view of the notorious experimental difficulties with accessing quark transversity distributions
directly through inclusive dilepton production with transversely polarized beam and target
\cite{RSPax},
several detour approaches were proposed in the present day literature.
Our strategy deals with the description of  hard
exclusive reactions within the collinear factorization approach in terms of transversity dependent
generalized parton distributions (GPDs). The chiral-oddity of the quark helicity flip operator
prevents the corresponding GPDs for contributing into photon or meson leptoproduction
amplitudes at the leading twist.
However, recently there were some attempts to circumvent this restriction
\cite{Ivanov:2002jj,Goloskokov:2013mba}.
In particular, the  transverse target spin asymmetries measured by COMPASS in vector meson exclusive leptoproduction
\cite{Adolph:2013zaa}
have been interpreted
\cite{Goloskokov:2013mba}
as a signal for transversity quark contributions.
On the contrary, the gluon sector  does not suffer from any selection rules and gluon helicity flip GPDs appear at  the leading twist level in
amplitudes of various hard exclusive reactions, for instance in the deeply virtual Compton
scattering (DVCS)
$O(\alpha_s)$
contribution to the leptoproduction of a real photon.
This contribution can be separated through a harmonic analysis
\cite{Diehl:1997bu}.
Particularly, the
$ (3\phi)$
modulation of the interference contribution to the unpolarized beam-longitudinally polarized target asymmetry seen in the HERMES data
\cite{Airapetian:2010aa}
for the DVCS on a nucleon may call for a significant gluon transversity contribution.

The practical applications of quark and gluon helicity flip GPD formalism require
the construction of flexible phenomenological parametrizations of these non-perturbative objects.
Below we review the main points of Ref.~\cite{Pire:2014fwa}
in which we established the crossed channel properties of parton helicity flip GPDs that are of
direct importance for a theoretically consistent model building  in the spirit of the double partial
wave expansion of GPDs. The phenomenological side of this study is postponed for future publications.

Let us briefly specify our set of conventions. Throughout this paper we  employ the usual
GPD kinematical notations for the average momentum
$P=\frac{1}{2}(p+p')$,
$t$-channel momentum transfer
$\Delta=p'-p$
and the skewness variable
$\xi$.
To the leading twist accuracy, the form factor decomposition of the non-forward nucleon matrix element of
the quark tensor operator
$\hat{O}_T^q$%
\footnote{Here and in eq. (\ref{Diehl_def}) we omit the Wilson gauge links by sticking to the light-cone gauge $A^+=0$. }
contracted with the appropriate projector
involves
$4$
invariant functions
\cite{Diehl:2001pm}:
\be
&&
\frac{1}{2}
\int
\frac{d \lambda}{2 \pi}
e^{i x P^+  \lambda }
\langle N(p' ) |
\bar{\Psi}(- \lambda n/2)
i \sigma^{+i}
\Psi ( \lambda n/2)
| N(p  ) \rangle 
= \frac{1}{2 P^+}
\bar{U}(p')
\left[
H^q_T i \sigma^{+i}+
\tilde{H}_T^q \frac{P^+ \Delta^i- \Delta^+ P^i}{m^2} \right.
\nonumber \\ &&
\left.
+E_T^q \frac{\gamma^+ \Delta^i- \Delta^+ \gamma^i}{2m}+
\tilde{E}_T^q \frac{\gamma^+ P^i- P^+ \gamma^i}{m}
\right] U(p),
\label{def_GPDs_T}
\ee
where $n$ denotes the light-cone vector and the Latin index $i=1,\,2$ refers to the transverse
spatial directions.
Each of the
four
invariant functions
$H^q_T$,
$\tilde{H}_T^q$,
$E^q_T$
and
$\tilde{E}_T^q$
depend on the  usual GPD variables.
Due to  hermiticity and  time reversal invariance, the
four
invariant functions are real valued. Moreover, one may check
\cite{Diehl:2001pm}
that
$H^q_T$, $\tilde{H}_T^q$, $E^q_T$
are even functions of
$\xi$
while
$\tilde{E}_T^q$
is an odd function of
$\xi$.

Similarly, the parametrization of the nucleon matrix element of the appropriately projected gluon tensor operator
$\hat{O}_T^g$
to the leading twist accuracy involves four
invariant functions
\cite{Diehl:2001pm}:
\be
&&
\frac{1}{P^+} \int \frac{d \lambda }{2 \pi} e^{i x P^+ \lambda}
\langle p'|  {\mathbb{S}} G^{+ i} (- \lambda n/2) G^{j + } ( \lambda n/2) | p \rangle 
= {\mathbb{S}} \; \frac{1}{2 P^+} \frac{P^+ \Delta^j-\Delta^+ P^j}{2 m P^+} \bar{U}(p') \Big[ H_T^g i \sigma^{+i}
\nonumber
\\ &&
+ \tilde{H}_T^g  \frac{P^+ \Delta^i-\Delta^+ P^i}{m^2}
+E_T^g  \frac{\gamma^+ \Delta^i-\Delta^+ \gamma^i}{2m}+
\tilde{E}_T^g  \frac{\gamma^+ P^i-P^+ \gamma^i}{m} \Big] U(p),
\label{Diehl_def}
\ee
where the
$\mathbb{S}$
symbol stands for the symmetrization in the two transverse spatial indices and removal of the corresponding trace.
From the combination of hermiticity and
$T$-invariance they are real valued.
$H_T^g$, $E_T^g$, $\tilde{H}_T^g$
are even functions of
$\xi$
while
$\tilde{E}_T^g$
is an odd function of
$\xi$.
Moreover, the $C$-invariance demands that
$H_T^g$, $E_T^g$, $\tilde{H}_T^g$
and
$\tilde{E}_T^g$
are  even functions of
$x$.

\section{SO$(3)$ partial wave expansion of quark and gluon GPDs with helicity flip}

The realistic strategy for extracting GPDs from the data relies on employing of phenomenologically
motivated GPD representations and simultaneous fitting procedures for the complete set of observable quantities.
The clue for building up a valid phenomenological representation for GPDs is provided by implementation of the
non-trivial requirements following from the fundamental properties of the underlying
quantum field theory.

Historically, one of the first parametrizations for GPDs suitable for phenomenological applications -
the famous Radyushkin double distribution Ansatz  - was based on the double distribution representation for GPDs.
It is employed within the extremely popular
Vanderhaeghen-Guichon-Guidal (VGG) code
\cite{Guidal:2010ig}
for the DVCS observables and saw some success in the description of the
available data.
The alternative way for building up of a GPD representation resides on the expansion of GPDs over a suitable
orthogonal polynomial basis in order to achieve the factorization of certain variable dependence. The appealing
possibility is to perform the expansion of GPDs over the conformal PW basis in order to achieve the
diagonalization of the leading order evolution operator.
Nowadays two main versions of such GPD representations are utilized in phenomenology:
the one based on the Mellin-Barnes integral techniques
\cite{Mueller:2005ed}
and the one using the idea of the Shuvaev-Noritzsch transformation
\cite{Shuvaev:1999fm}.
It turns out
extremely instructive to further expand the conformal moments over the basis of
the $t$-channel
${\rm SO}(3)$
rotation group partial
waves. In the context of the Shuvaev transform techniques the resulting GPD representation
is known as the dual parametrization of GPDs \cite{Polyakov:2002wz}.
Within the Mellin-Barnes integral techniques  \cite{Mueller:2005ed} this version of the conformal partial wave expansion
is referred in the literature as the
${\rm SO}(3)$
partial wave expansion.
Each version of the formalism employs a rather intricate mathematical apparatus, however,
as argued in
\cite{MuellKS},
these two approaches turn out to be completely equivalent.

Finding out the combinations of GPDs suitable for the $t$-channel
${\rm SO}(3)$
partial waves and the choice of the appropriate basis of the orthogonal polynomials represents an important task.
For example, for the case of the unpolarized quark and gluon nucleon GPDs this kind of analysis gives rise to the
so-called electric and magnetic combinations of GPDs
\cite{Diehl}:
$
H^{E \, \{q,g\}}=H^{\{q,g\}}+ \tau E^{\{q,g\}}; \ \ \
H^{M \,\{q,g\}}=H^{\{q,g\}}+ E^{\{q,g\}},
$
where
$
\tau \equiv \frac{\Delta^2}{4 m^2}.
$
These combinations are to be expanded respectively in terms of
$ P_J(\cos \theta)
$
and
$ P'_J(\cos \theta)$,
with
$P_J(\cos \theta)$
standing for the Legendre polynomials and
$\theta$
referring to the
$t$-channel scattering angle in the
$N \bar{N}$
center-of-mass frame.

Following the receipt of sect. 4.2 of
ref.~\cite{Diehl},
in order to identify  the combinations of quark helicity flip  GPDs suitable
for the partial wave expansion in the $t$-channel partial waves, we
consider the form factor decomposition of the $N$-th Mellin moments
of quark and gluon helicity flip  GPDs analytically continued to the cross channel
($t>0$).
Thus we are dealing with the form factor decomposition of the
$N$-th
Mellin moments  of  quark helicity flip
$N \bar{N}$
generalized distribution amplitudes (GDAs).
To find which partial waves contribute into the  corresponding matrix elements,
we compute the spin-tensor structures for spinors of definite (usual) helicity in the
$N \bar{N}$
center-of-mass (CMS) frame using the explicit expressions
for the nucleon spinors  with definite ordinary helicity.
Let us briefly review the main stages of the calculation for the case of quark helicity flip GPDs.
We project out the combination of the
matrix elements with definite helicity
$J_3=\pm 1$ of the corresponding operator:
\be
&&
\langle  N(p',\lambda') \bar{N}( -p,\lambda)| \hat{O}_{ T}^{q \, + 1, \, ++ \ldots + }|0 \rangle
\pm i \langle  N(p',\lambda') \bar{N}( -p,\lambda)| \hat{O}_{  T}^{q\, + 2, \, ++ \ldots + }|0\rangle
\nonumber \\ &&
\equiv \langle  N(p',\lambda') \bar{N}( -p,\lambda)| \hat{O}_{  T}^{q \, + (1 \pm i2), \, ++ \ldots + }|0 \rangle.
\label{helicity_combinations_quarks}
\ee
The combinations
(\ref{helicity_combinations_quarks})
possess definite phases depending on the azimuthal angle
$\phi$.
Now one can decompose
(\ref{helicity_combinations_quarks})
in the partial waves with total angular momentum
$J$.
The
$\theta$
dependence is governed by the Wigner ``small-$d$'' rotation functions
 $d^J_{J^3, |\lambda'-\lambda|}$.
For the case
$|\lambda'-\lambda|=0$ ({\it i.e.} the aligned configuration of nucleon and antinucleon helicities%
\footnote{Note, that our hadron helicity labeling refers to the $t$-channel. Obviously,
when crossing back to the direct channel the helicity
$\lambda$
is reversed. }
($N^\uparrow \bar{N}^\uparrow$ or $N^\downarrow \bar{N}^\downarrow$))
and
$J_3= \pm 1$
one has to use
\be
d^J_{\pm 1, 0}(\theta) = (\pm 1) \frac{1}{\sqrt{J(J+1)} } \sin \theta \, P'_J(\cos \theta).
\ee

For the case when
$|\lambda'-\lambda|=1$
({\it i.e.} the opposite  helicity configuration of nucleon and antinucleon:
$N^\uparrow \bar{N}^\downarrow$ or $N^\downarrow \bar{N}^\uparrow$) depending on
the operator helicity
$J_3= \pm 1$
(\ref{helicity_combinations_quarks})
is to be expanded in
\be
d^J_{\pm 1, 1}(\theta)=   \frac{1}{J(J+1)} (1 \pm \cos \theta) \big[ P'_J(\cos \theta) + \cos \theta P''_J(\cos \theta) \mp P''_J(\cos \theta) \big].
\ee

After the inverse crossing
(\ref{helicity_combinations_quarks})
back to the $s$-channel, within the DVCS kinematics
$\cos \theta$
up to higher twist corrections becomes
$
\cos \theta \rightarrow \frac{1}{\xi \beta}+ O(1/Q^2),
$
where $\beta \equiv \sqrt{1- \frac{4m^2}{t}}$.
At this stage we switch to massless hadrons so that we could consider
hadron helicities as true quantum numbers thus making simple
the crossing relation between the corresponding partial amplitudes (in particular excluding mixing).
This implies setting
$\beta=1$
(which means systematically neglecting the threshold corrections
$\sim \sqrt{1-\frac{4m^2}{t}}$).
However, up to the very end  we keep the non-zero mass within the Dirac spinors to keep the counting of independent
tensor structures.
Finally, we conclude that the following combinations of quark helicity flip GPDs are to be expanded in
$P'_J(1/\xi)$:
\be
&&
\tau
\tilde{H}_T^q(x,\xi,\Delta^2) -\frac{1}{2} E_T^q(x,\xi,\Delta^2); \nonumber \\ &&
-  H_T^q(x,\xi,\Delta^2)+
\tau
\tilde{H}_T^q(x,\xi,\Delta^2) -\frac{1}{2} E_T^q(x,\xi,\Delta^2).
\label{combi}
\ee
while the combinations
\be
H_T^q(x,\xi,\Delta^2)+
\tau \tilde{H}_T^q(x,\xi,\Delta^2)  \pm
\tau \tilde{E}_T^q(x,\xi,\Delta^2)
\label{comb_without_definite}
\ee
are to be expanded in
$
 P'_J(1/\xi) +   \frac{1 \mp \xi}{\xi}  P''_J(1/\xi).
$

The gluon case can be considered according to the same pattern (see Sec.3.2 of Ref.~\cite{Pire:2014fwa})
giving rise to the combinations of matrix elements expanded in terms of the Wigner functions
$d_{\pm2,0}^J (\theta)$, $d_{\pm2,1}^J (\theta)$ and  $d_{0,0}^J (\theta)$.

The method also allows to work out the set of selection rules for the
$J^{PC}$
quantum numbers for the $t$-channel resonance exchanges
contributing into the $N$-th Mellin moments of quark helicity flip GPDs.
Due to the
$CPT$
invariance, this kind of
$J^{PC}$
matching in the cross channel automatically ensures the
$T$
invariance and the correct counting of the independent generalized form factors of
the operator matrix element in the direct channel.
The  selection rules we establish coincide with those
worked out with the general method of X.~Ji and R.~Lebed \cite{Ji:2000id}.

The alternative method to work out the set of quark and gluon helicity flip GPDs suitable for the
partial wave expansion in the cross-channel partial waves consists in the explicit calculation of
the cross channel spin-$J$ resonance contributions into corresponding GPD.
The advantage of this method is that it is fully covariant and allows to determine the net
resonance exchange contributions into scalar invariant functions
$H_{T}^{q,g}$,
$E_{T}^{q,g}$,
$\tilde{H}_{T}^{q,g}$,
$\tilde{E}_{T}^{q,g}$.
Within this approach
the nucleon matrix element of the light-cone operator
$\hat{O}$
is represented as an infinite sum of  $t$-channel resonance exchange contributions.
Symbolically it can be written in the following form:
\be
&&
\langle N(p')|\,
\hat{O} \,
| N(p) \rangle
\sim
\sum_{R_J}
\sum_{{\rm polarizations} \atop {\rm of \;} R_J }
\frac{1}{\Delta^2-M_{R_J}^2}
\times
\underbrace{\; \langle N(p') R_J(\Delta) | N(p)  \rangle \; }_{V_{R_J N  N}  {\rm \; eff. \, vertex} } \; \otimes
\underbrace{ \langle 0 |
\hat{O}
|  R_J(\Delta)\rangle }_{ {\rm Fourier \, Transform\, of \, DA \, of \,} R_J}, \nonumber
\label{Ma trix_element}
\ee
where
$M_{R_J}$ stand for the resonance masses and
$\otimes$ denotes the convolution in the appropriate Lorentz indices. The resulting on-shell polarization sums for spin-$J$ resonances can be performed with the contracted projectors method
(see {\it e.g.} Chapter I of
\cite{Alfaro_red_book}).
This calculation allows to recover the same combinations of quark and gluon helicity flip GPDs suitable for ${\rm SO}(3)$ PW expansion in the $t$-channel.
Moreover, as a byproduct, we build up a simple $f_2(1270)$ meson exchange model for gluon helicity flip GPDs that is similar to $b_1$ meson exchange
model for quark helicity flip GPDs suggested in
\cite{Enberg:2006he}.
The relevant $f_2 N \bar{N}$ coupling constants can be obtained from low energy $NN$ scattering studies and the gluon helicity flip
distribution amplitude normalization constant can be estimated as suggested in \cite{Braun:2000cs}.
This model  allows for the first time to work out the physical normalization of  gluon helicity flip GPDs.

\section{Conclusions}

In order to construct  a theoretically  consistent parametrization of  these hadronic matrix elements,
we work out the set of combinations of the transversity GPDs suitable for the
${\rm SO}(3)$
PW expansion in the cross-channel. This universal result  will help us to build up a flexible parametrization
of these important hadronic non-perturbative quantities, using for instance the approaches
based on the conformal PW expansion of GPDs such as the Mellin-Barnes integral or the dual parametrization techniques.
We also propose a simple  $f_2(1270)$ meson exchange model for gluon helicity flip GPDs that allows to estimate the
physical normalization of  gluon transversity effects.
This work is partly supported by the Polish Grant NCN
No DEC-2011/01/B/ST2/03915,  the Joint Research Activity ``Study of Strongly
Interacting Matter'' (acronym HadronPhysics3, Grant 283286) under the Seventh
Framework Programme of the European Community, by the COPIN-IN2P3 Agreement,
by the French grant ANR PARTONS (ANR-12-MONU-0008-01)
and by the Tournesol 2014 Wallonia-Brussels-France Cooperation Programme.

\end{document}